\documentclass[preprint,12pt,showpacs,aps,amsmath,amssymb,nofootinbib,superscriptaddress,hyperref,showkeys]{revtex4-1} 

\usepackage[utf8]{inputenc}
\usepackage{color}
\usepackage{epsfig} 
\usepackage{wasysym} 
\usepackage{mathrsfs} 
\usepackage{graphicx} 
\usepackage{amsfonts} 
\usepackage{amsbsy} 
\usepackage{amscd} 
\usepackage{pstricks} 
\usepackage{multirow}
\usepackage{slashed} 
\usepackage{tikz}
\usetikzlibrary{decorations.pathmorphing,decorations.markings}
\newcommand{\beq}{\begin{equation}}
\newcommand{\eeq}{\end{equation}}
\newcommand{\bea}{\begin{eqnarray}}
\newcommand{\eea}{\end{eqnarray}}
\newcommand{\beas}{\begin{eqnarray*}}
\newcommand{\eeas}{\end{eqnarray*}}
\newcommand{\bi}{\begin{itemize}}
\newcommand{\ei}{\end{itemize}}

\def\tev{\,{\ifmmode\mathrm {TeV}\else TeV\fi}}
\def\gev{\,{\ifmmode\mathrm {GeV}\else GeV\fi}}
\def\to{\rightarrow}

\allowdisplaybreaks

\begin{document}

\title{Constraining Sterile Neutrinos from Precision Higgs Data}

\author{Arindam Das\footnote{arindam@kias.re.kr}}
\affiliation{School of Physics, KIAS, Seoul 130-722, Korea}
\affiliation{Department of Physics \& Astronomy, Seoul National University 1 Gwanak-ro, Gwanak-gu, Seoul 08826, Korea}
\affiliation{Korea Neutrino Research Center, Bldg 23-312, Seoul National University, Sillim-dong, Gwanak-gu, Seoul 08826, Korea}
\author{P. S. Bhupal Dev\footnote{bdev@wustl.edu}}
\affiliation{Department of Physics and McDonnell Center for the Space Sciences, Washington University, St. Louis, MO 63130, USA}
\author{C. S. Kim\footnote{cskim@yonsei.ac.kr}}
\affiliation{Department of Physics and IPAP, Yonsei University, Seoul 120-749, Korea }

\preprint{\today}

\begin{abstract}
We use the LHC Higgs data to derive updated constraints on electroweak-scale sterile neutrinos that naturally occur in many low-scale seesaw extensions of the Standard Model to explain the neutrino masses.  We also analyze the signal sensitivity for a new final state involving a single charged lepton and two jets with missing energy, which arises from the decay of sterile neutrinos produced through the Higgs and $W,Z$ boson mediated processes at the LHC. Future prospects of these sterile neutrino signals in precision Higgs measurements, as well as at a future 100 TeV collider, are also discussed. 
%

\end{abstract}


\maketitle

\section{Introduction}
\label{sec:intro}
The results of the neutrino oscillation experiments~\cite{Olive:2016xmw} indicate tiny but non-zero masses for at least two active neutrinos, which is so far the only laboratory evidence for the existence  of beyond the Standard Model (SM) physics. A simple paradigm that naturally explains the smallness of neutrino masses is the so-called type-I seesaw~\cite{typeI-seesaw1, typeI-seesaw2, typeI-seesaw3, typeI-seesaw4, typeI-seesaw5, typeI-seesaw6}, which requires SM-singlet heavy Majorana neutrinos (generically denoted here by $N$). In the minimal version of the type-I seesaw, as well as its variants, such as inverse~\cite{inverse-seesaw1, inverse-seesaw2}, linear~\cite{Akhmedov:1995vm, Malinsky:2005bi} and generalized~\cite{Gavela:2009cd, Dev:2012sg} seesaw, the sterile neutrinos, being gauge-singlets, couple to the SM sector only through their mixing with the active neutrinos via Dirac Yukawa couplings (hence the name `sterile'). 

In a  bottom-up phenomenological approach, the seesaw scale is a priori unknown and can be anywhere between the eV scale and the grand unification scale~\cite{Mohapatra:2005wg, Drewes:2013gca}. In the Large Hadron Collider (LHC) era, a particularly interesting mass range for the sterile neutrinos is the sub-TeV scale, which is kinematically accessible at the LHC energies, thereby providing a unique opportunity to directly test the low-scale seesaw mechanism. The phenomenological aspects of the heavy sterile neutrino production at colliders have been widely discussed; see e.g. Refs.~\cite{Keung:1983uu,Datta:1992qw,Datta:1993nm,Almeida:2000pz,Panella:2001wq,Han:2006ip,Bray:2007ru,delAguila:2007qnc,
Huitu:2008gf,delAguila:2008cj,delAguila:2008hw,Atre:2009rg,Perez:2009mu,Chen:2011hc,Das:2012ze,
Han:2012vk,Chen:2013fna,Dev:2013wba,Deppisch:2013jxa,
Das:2014jxa,Bambhaniya:2014kga,Bambhaniya:2014hla,Alva:2014gxa,Antusch:2015mia,
Deppisch:2015qwa,Banerjee:2015gca,Alekhin:2015byh,Gluza:2015goa,Ng:2015hba,
Dev:2015pga,Peng:2015haa,Asaka:2015oia,Dev:2015kca,Das:2015toa,
Leonardi:2015qna,Antusch:2015gjw,Kang:2015uoc,Dev:2016vif,Das:2016akd, Lindner:2016lxq,Degrande:2016aje,Das:2016hof,
Hessler:2014ssa,
Gluza:2016qqv,Mondal:2016kof,Golling:2016gvc,Fischer:2016rsh,Anamiati:2016uxp,
Duarte:2016caz,Antusch:2016ejd,Agashe:2016ttz,Dev:2016gvv,Guo:2017ybk,Das:2017pvt,
Biswal:2017nfl,Das:2017nvm} and references therein. The latest experimental search results of the sterile neutrinos in the `smoking gun' same-sign dilepton channel at the LHC can be found in  Refs.~\cite{ATLAS8, CMS8, Khachatryan:2016olu}.

The success of the sterile neutrino searches at colliders in the same-sign dilepton channel crucially depends on both the Majorana nature of the sterile neutrinos, as well as the size of the active-sterile neutrino mixing parameter, in the minimal seesaw scenario. In the canonical type-I seesaw, one expects the mixing parameter $V_{\ell N}\simeq M_DM_N^{-1}\lesssim 10^{-6} \sqrt{(100~{\rm GeV})/M_N}$, where $M_D$ and $M_N$ are respectively the Dirac and Majorana masses in the seesaw matrix. Possible cancellations in the seesaw matrix could allow for a larger mixing parameter even for TeV-scale $M_N$~\cite{Pilaftsis:1991ug,Gluza:2002vs,Kersten:2007vk,Xing:2009in,He:2009ua,Adhikari:2010yt,Ibarra:2010xw,
Deppisch:2010fr,Mitra:2011qr,Dev:2013oxa,Haba:2016lxc}, justifying the direct collider searches. However, most of these scenarios lead to a suppressed lepton number violation, mainly due to the stringent constraints from neutrino oscillation data and neutrinoless double beta decay ($0\nu\beta\beta$)~\cite{Kersten:2007vk, Ibarra:2010xw, Lopez-Pavon:2015cga, Drewes:2016jae}. Therefore, it is important to also look for the signals that are not suppressed by the effective lepton number violation in the theory, i.e. applicable regardless of the Majorana nature of the sterile neutrinos. Some examples are opposite-sign dilepton~\cite{Gluza:2015goa, Dev:2015pga, Gluza:2016qqv, Anamiati:2016uxp, CMS8} and trilepton~\cite{Chen:2011hc, Das:2014jxa} signals. For sub-electroweak scale sterile neutrinos, there are additional collider signals of this kind, such as displaced vertices~\cite{Gago:2015vma, Helo:2013esa, Dev:2016vle, Dev:2017dui, Deppisch:2013cya, Accomando:2016sge, Accomando:2016rpc, Antusch:2016vyf}, decays of $W$-boson~\cite{Izaguirre:2015pga, Dib:2015oka, Dib:2016wge, Blondel:2014bra, Dib:2017iva, Dib:2017vux} and decays of SM Higgs~\cite{Dev:2012zg, Cely:2012bz, Dermisek:2014qca, Das:2017rsu} which are complementary to the direct searches~\cite{ATLAS8, CMS8, Khachatryan:2016olu}.

In this paper we revisit the sterile neutrino production via the SM Higgs decay in light of the current and future precision Higgs measurements. In particular, the Dirac Yukawa coupling responsible for the active-sterile neutrino mixing and the active neutrino mass also induces the anomalous Higgs decay $h\to \nu N$, if kinematically allowed. This has two potentially observable effects on the SM Higgs properties: (i) enhancement of the total Higgs decay width, as compared to its SM predicted value, and (ii) enhancement of the Higgs signal strength in certain channels, depending on the sterile neutrino decay, which in turn leads to a suppression of the Higgs signal strength in the other SM channels. Therefore, precision  measurements of the Higgs boson properties could yield important constraints on the sterile neutrino mass and mixing parameters. 

We illustrate this effect by analyzing the Higgs boson production and decay at the LHC, followed by the sterile neutrino decay to a charged lepton and $W$ boson, which mimics the SM $h\to WW^*$ channel. So using  the $\sqrt s=8$ TeV LHC data in the $h\to WW^*$ search channel, which is largely consistent with the SM expectations, we derive constraints on the active-sterile neutrino mixing parameter $V_{\ell N}$ as a function of the sterile neutrino mass. Based on this analysis, we also make conservative predictions for the future limits at the $\sqrt s=14$ TeV high-luminosity (HL) LHC, as well as a futuristic $\sqrt s=100$ TeV hadron collider, such as FCC-hh or SPPC. We find that our limits could be comparable to, or in some cases, better than the current best limits for sterile neutrino masses in the vicinity of the Higgs boson mass. Our study includes two possibilities for the $W$ decay, namely, (i) leptonic mode leading to $2\ell 2\nu$ final state, and  (ii) hadronic mode leading to $\ell \nu jj$ final state. We find that the leptonic mode has better sensitivity at the LHC, mainly due to the smaller background, as compared to the hadronic decay channel. 

The rest of the paper is organized as follows: in Section~\ref{sec:ndecay} we review the decay modes of the sterile neutrino both above and below the SM gauge boson mass scales. In Section~\ref{sec:bounds}, we discuss the sterile neutrino production in SM Higgs boson decay and analyze the resultant $2\ell 2\nu$ final state to derive constraints on the sterile neutrino parameter space. In Section~\ref{sec:calc}, we analyze a new final state from the sterile neutrino production, namely, the $\ell \nu jj$ channel and its discovery prospects at $\sqrt s=14$ and 100 TeV hadron colliders. Our conclusions are given in Section~\ref{sec:conc}.  
\section{Sterile neutrino decay}
\label{sec:ndecay}
We consider the minimal singlet seesaw extension of the SM, where the production and decay properties of the sterile neutrino are governed by its mass and mixing with the active neutrinos. We do not want to go into the specific details of neutrino mass models, but keep our discussion generic, regardless of whether the sterile neutrinos are Majorana or pseudo-Dirac particles. In this sense, our results are applicable to all low-scale singlet seesaw models with the SM gauge group, including the minimal type-I seesaw~\cite{typeI-seesaw1, typeI-seesaw2, typeI-seesaw3, typeI-seesaw4, typeI-seesaw5, typeI-seesaw6}, as well as its variants, such as inverse~\cite{inverse-seesaw1, inverse-seesaw2}, linear~\cite{Akhmedov:1995vm, Malinsky:2005bi} and generalized~\cite{Gavela:2009cd, Dev:2012sg} seesaw.  

Due to the active-sterile neutrino mixing, a light neutrino flavor eigenstate ($\nu_\ell$) is a linear combination of the light ($\nu_m$) and heavy ($N_m$) neutrino mass eigenstates:    
\bea 
  \nu_\ell \ \simeq \ U_{\ell m} \nu_m  + V_{\ell n} N_n \, ,  
\eea 
where $U$ is the $3\times 3$ light neutrino mixing matrix (which is same as the PMNS mixing matrix to leading order, if we ignore the non-unitarity effects), and $V\simeq M_DM_N^{-1}$  is the active-sterile mixing parameter. The charged-current (CC) interaction in the lepton sector is then given by  
\bea 
\mathcal{L}_{\rm CC} \ = \ 
 -\frac{g}{\sqrt{2}} W_{\mu}
  \bar{\ell} \gamma^{\mu} P_L 
   \left[ U_{\ell m} \nu_m+  V_{\ell n} N_n \right] + {\rm H. c.}, 
\label{CC}
\eea
where $g$ is the $SU(2)_L$ gauge coupling and  $P_L =(1- \gamma_5)/2$ is the left-chiral projection operator. 
Similarly, the neutral-current (NC) interaction is given by 
\bea 
\mathcal{L}_{\rm NC} & \ = \ & -\frac{g}{2 \cos\theta_w}  Z_{\mu} \left[ (U^\dag U)_{mn} \bar{\nu}_m \gamma^{\mu} P_L \nu_n + (U^\dag V)_{mn} \bar{\nu}_m\gamma^\mu P_L N_n + (V^\dag V)_{mn}\bar{N}_m\gamma^\mu P_L N_n\right]  \nonumber \\
&& \hspace{5cm}+ {\rm H. c.} , 
\label{NC}
\eea
 where $\theta_w$ is the weak mixing angle. Thus, the interactions of the sterile neutrino with the SM gauge sector are all suppressed by powers of the mixing matrix $V$. 

Similarly, the relevant Yukawa interaction is given by  
\bea
\mathcal{L}_{Y} \ \supset \ -Y_{D_{\ell m}} \bar{L}_\ell \phi N_m + {\rm H.c.} \, ,
\label{yuk}
\eea
where $L$ and $\phi$ are the $SU(2)_L$ lepton and Higgs doublets, respectively. After electroweak (EW) symmetry breaking by the vacuum expectation value (VEV) of the Higgs doublet, $\langle \phi^0 \rangle =v$, we get the Dirac mass term $M_D= vY_D$. So the Yukawa coupling of the sterile neutrino to the SM Higgs is given by $Y_D=VM_N/v$, which is also suppressed by $V$.     
 
For simplicity, we will assume that only the lightest heavy neutrino mass eigenstate (denoted here simply by $N$) is kinematically accessible at colliders, and denote the corresponding mixing parameter as simply $V_{\ell N}$, which is the only free parameter in our phenomenological analysis, apart from the sterile neutrino mass $M_N$. From Eqs.~\eqref{CC}, \eqref{NC} and \eqref{yuk}, we see that there are three decay modes for the sterile neutrino, if kinematically allowed: 
 $N \to \ell^- W^+$, $\nu_\ell Z$, $\nu_\ell h$, where $h$ is the SM Higgs boson (the only physical scalar remnant of the doublet $\phi$). The corresponding partial decay widths are respectively given by
\bea
\Gamma(N \rightarrow \ell^- W^+) 
 & \ = \ & \frac{g^2 |V_{\ell N}|^{2}}{64 \pi}\frac{M_N^3}{M_W^2}\left(1-\frac{M_W^2}{M_N^2}\right)^2\left(1+\frac{2M_W^2}{M_N^2}\right), \\
\Gamma(N \rightarrow \nu_\ell Z) 
 & \ = \ & \frac{g^2 |V_{\ell N}|^{2}}{128 \pi} 
\frac{M_N^3}{M_W^2}\left(1-\frac{M_Z^2}{M_N^2}\right)^2\left(1+\frac{2M_Z^2}{M_N^2}\right),
\\
\Gamma(N_1 \rightarrow \nu_\ell h) 
 &=& \frac{|V_{\ell N}|^{2}}{128\pi} \frac{M_N^3}{M_W^2}\left(1-\frac{M_h^2}{M_N^2}\right)^2.
\label{widths}
\eea 
The total decay width is just the sum of the above three partial widths for each flavor and summed over all lepton flavors. If $N$ is a Majorana particle, the charge-conjugate 
modes, namely, $\ell^+ W^-$, $\bar{\nu} Z$ and $\bar{\nu} h$ are also allowed, so there is an additional factor of 2.

For $M_N<M_W$, none of these two-body decay modes are kinematically allowed. In this case, the sterile neutrino will have three-body decays dominantly mediated by the SM gauge bosons. 
 The corresponding partial decay widths when the off-shell SM gauge bosons decay leptonically,  are given by
 \bea
 \Gamma(N \rightarrow \ell_1^{-} \ell_2^{+} \nu_{\ell_2}) & \ \simeq \ & \frac{|V_{\ell_1 N}|^{2} G_{F}^{2} M_N^{5}}{192 \pi^{3}},  \\
 \Gamma(N \rightarrow \nu_{\ell_1} \ell_2^{+} \ell_2^{-}) & \ \simeq \ & \frac{|V_{\ell_1 N}|^{2} G_{F}^{2} M_N^{5}}{96 \pi^{3}} \big(g_{L} g_{R}+ g_{L}^{2}+ g_{R}^{2}\big),  \\
 \Gamma(N \rightarrow \nu_{\ell} \ell^{+} \ell^{-}) & \ \simeq \ & \frac{|V_{\ell N}|^{2} G_{F}^{2} M_N^{5}}{96 \pi^{3}} \big(g_{L} g_{R}+ g_{L}^{2}+ g_{R}^{2}+ 1+ 2 g_{L}\big), \\
 \Gamma(N \rightarrow \nu_{\ell_1} \nu_{\ell_2} \bar{\nu}_{\ell_2}) & \ \simeq \ & \frac{|V_{\ell_1 N}|^{2} G_{F}^{2} M_N^{5}}{96 \pi^{3}} , 
 \label{lep}
  \eea
  and the corresponding decay widths when the SM gauge bosons decay hadronically are given by 
 \bea
 \Gamma(N \rightarrow \ell^{-} j j) &\ \simeq \ & 3\frac{|V_{\ell N}|^{2} G_{F}^{2} M_N^{5}}{192 \pi^{3}}, \label{had0} \\
 \Gamma(N \rightarrow \nu_{\ell} jj) & \ \simeq \ & 3\frac{|V_{\ell N}|^{2} G_{F}^{2} M_N^{5}}{96 \pi^{3}} \big(g_{L} g_{R}+ g_{L}^{2}+ g_{R}^{2}\big),
 \label{had}
  \eea
where $g_{L}= -\frac{1}{2}+ \sin^2\theta_{w}$, $g_{R}= \sin^2\theta_{w}$, and the factor 3 in Eqs.~\eqref{had0} and \eqref{had} is the color factor.    
Thus the total decay width for the sterile neutrino with $M_{N} < M_{W}$  is given by 
 \bea
\Gamma_{N}& \ \simeq \ & 3\big[2\Gamma(N \rightarrow e^{-} \mu^{+} \nu_{\mu})+ 2 \Gamma(N \rightarrow \nu_{e} \mu^{+} \mu^{-})+  \Gamma(N \rightarrow \nu_{\mu} \mu^{+} \mu^{-}) \nonumber \\
 & + & \Gamma(N \rightarrow \nu_{e} \nu_{\mu} \nu_{\mu})+ 2 \Gamma(N \rightarrow e^{-} j j)+ 5  \Gamma(N \rightarrow \nu_{e} jj)\big] \, .
 \label{totdecay}
\eea 
In Eq.~\eqref{totdecay} the factor 2 in the first two terms is due to the two flavors $\ell_2\neq \ell_1$, whereas the third one is fixed by the heavy neutrino vertex. The factor of 2 in front of the fifth term is taken for $ud$ and $cs$ pairs.
The factor of 5 in front of the sixth term is introduced for $uu$, $dd$, $ss$, $cc$ and $bb$ pairs. The overall factor of 3 is for the sum over three lepton flavors. Here we have neglected the lepton masses. For more exact expressions, see e.g. Ref.~\cite{Helo:2013esa}. 

%
 %
\section{Sterile neutrino production from Higgs Decay}
\label{sec:bounds}
\begin{figure}
\begin{center}
\includegraphics[scale=1]{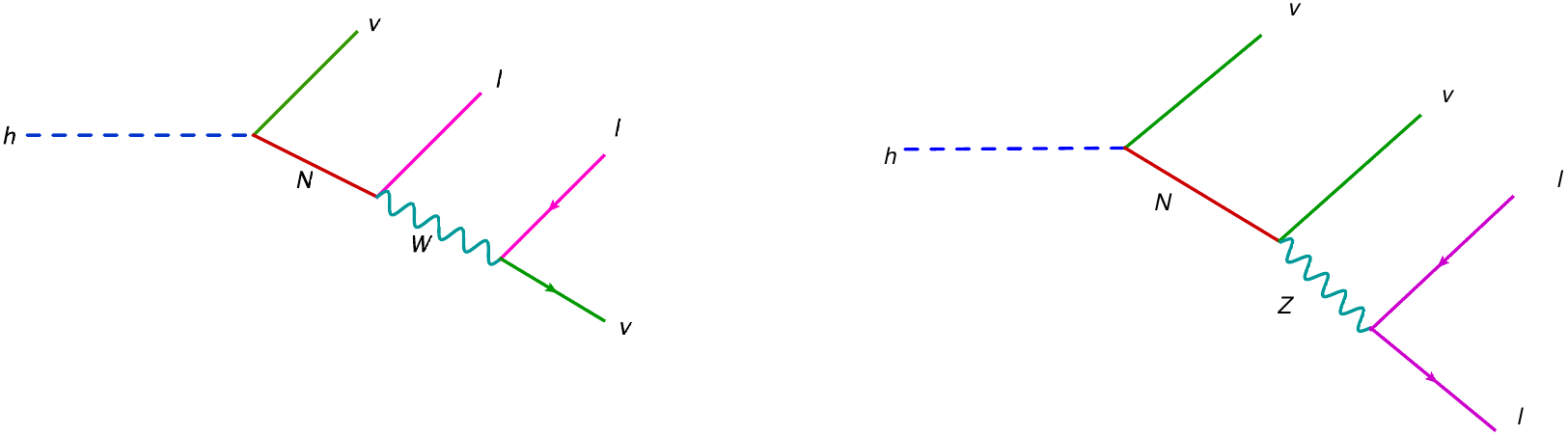}
\end{center}
\caption{Higgs decay to sterile neutrino giving rise to $2\ell2\nu$ final state.}
\label{fig1}
\end{figure}
The new Yukawa interaction in Eq.~\eqref{yuk} gives rise to a new decay mode for the SM Higgs, $h\to N\nu$, if kinematically allowed. Depending on the $N$ decay, we will have different final states. In this section, we will examine the leptonic final states $2\ell 2\nu$, which can arise from either  $N\to \ell^-_1W^{+(*)} \to \ell^-_1\ell_2^+ \nu$ (with both $\ell_1=\ell_2$ and $\ell_1\neq \ell_2$ possibilities) or $N\to \nu Z^{(*)} \to \nu \ell^-\ell^+$. The corresponding 
Feynman diagrams are given in Fig.~\ref{fig1}. The important thing to note here is that these final states mimics the SM process $h\to WW^*\to 2\ell 2\nu$, and therefore, enhance the $h\to WW^*$ signal strength~\cite{Dev:2012zg}, while suppress the other SM decay modes, with respect to the SM predictions. It is worth mentioning here that the $h\to WW^*$ channel has the second largest branching fraction (22\%) in the SM for $M_h=125$ GeV and is a good candidate for studying Higgs boson properties.   

Before going into the experimental details, we would like to point out that due to the new Yukawa interaction in Eq.~\eqref{yuk}, the total decay width of the Higgs boson is also enhanced with respect to its SM predicted value: 
\bea
\Gamma_h \ = \ \Gamma_{\rm SM}+ \Gamma_{\rm new}
\label{gtot}
\eea
where $\Gamma_{\rm SM}\simeq 4.1$ MeV for $M_h=125$ GeV~\cite{Heinemeyer:2013tqa} and 
\bea
\Gamma_{\rm new} \ = \ \frac{Y_{D}^2 M_{h}}{8 \pi} \left(1-\frac{M_{N}^{2}}{M_{h}^{2}}\right)^2
\label{h2N}
\eea
From the LHC studies of Higgs boson off-shell production in gluon fusion and vector boson fusion processes, an upper limit on the total width of the SM Higgs boson of $\Gamma_h<13$ MeV at 90\% CL has been derived~\cite{Khachatryan:2016ctc}. From Eq.~\eqref{gtot}, this implies an upper limit on the Yukawa coupling, and hence, on the mixing parameter $|V_{\ell N}|^2$. This is shown by the red solid curves in Figure~\ref{fig2} (all panels). With future precision Higgs measurements, this limit could be further improved. For instance, up to 10\% precision in Higgs total width can be achieved at a 100 TeV $pp$ collider: $\Gamma_h<1.1\Gamma_{\rm SM}$~\cite{Contino:2016spe}, which corresponds to a limit on the mixing parameter as shown by the red dashed curve in Figure~\ref{fig2}. A future lepton collider can achieve an accuracy of up to 5\%~\cite{Han:2013kya} (2.5\% with the luminosity upgrade~\cite{Durig:2014lfa}). 

We can obtain a better limit on the mixing parameter by analyzing the LHC Higgs data in the $h\to WW^*\to 2\ell 2\nu$ channel, which are largely consistent with the SM predictions and do not allow a significant deviation. The experimental analyses in this channel have been performed by both CMS and ATLAS with full $\sqrt s=8$~\cite{Chatrchyan:2013iaa, ATLAS:2014aga} and early 13 TeV LHC datasets~\cite{CMS:2016nfx, ATLAS:2016kjy}. For concreteness, we will reinterpret the cut-based analysis presented in Ref.~\cite{ATLAS:2014aga} to extract an upper bound on the extra contribution from $h\to \nu N\to 2\ell 2\nu$.\footnote{One can also use the $h\to ZZ^*\to 2\ell 2\nu$ channel~\cite{Khachatryan:2015pba, Aaboud:2016urj} to derive similar constraints. } 

For this we implement our model in the event generator  {\tt MadGraph5-aMC@NLO} \cite{aMC}.  
The showering and hadronization of the events were performed with {\tt PYTHIA6.4} ~\cite{Pyth} bundled in {\tt MadGraph} with {\tt anti-$k_{T}$} algorithm, while the jets are 
clustered using {\tt FastJet} simulation~\cite{FJ}. The hadronic cross-sections have been calculated using the {\tt CTEQ6L1} parton distribution functions (PDF)~\cite{Dulat:2015mca}. We use the hadronized events in {\tt Delphes}~\cite{Delphes} to simulate the detector response. The event selection criteria are chosen following the cut-based analysis in Ref.~\cite{ATLAS:2014aga}.  

In our analysis, we have four different mass regions for the heavy neutrino, as  given in Table~\ref{tab1}.
\begin{table}[t!]
\begin{center}
\begin{tabular}{c| c }
\hline
Region & Mass range \\ \hline
$1$ & $M_N <  M_W$ \\
$2$ & $M_W < M_N < M_Z$ \\
$3$ & $M_Z < M_N < M_h$\\
$4$ & $M_N > M_h$\\
\hline
\end{tabular}
\end{center}
\caption{Four different mass regions of the heavy neutrino considered in our analysis.}
\label{tab1}
\end{table}
When $M_N < M_W$ (region 1), the produced heavy neutrino will have three-body decays to $\ell_{1} \bar{\ell}_{1} \nu$ (mediated by both $W$ and $Z$ bosons), $\ell_{1} \bar{\ell}_{2} \nu$ (mediated by $W$), and $\nu \ell_{2} \bar{\ell}_{2}$ (mediated by $Z$). When  $M_W<M_N< M_Z$ (region 2),  the three-body decay of the heavy neutrino will contribute to $\nu \ell_{1} \bar{\ell}_{1} $ and $\nu \ell_{2} \bar{\ell}_{2}$ (mediated by the $Z$ boson), whereas the $W$-boson mediated process $N\to \ell_1 W\to \ell_1\ell_2 \nu$ is a two-body decay. Similarly, when $M_N < M_h$, the Higgs boson decays into on-shell $N\nu$ through the Dirac Yukawa coupling given in Eq.~\eqref{yuk}. On the other hand, for $M_N > M_{h}$, the heavy neutrino behaves as an intermediate-state propagator in the process $pp\to h\to \nu N\to 2\ell 2\nu$.

In this analysis, we have three types of events for the $\ell\bar{\ell}\nu\bar{\nu}$ depending upon the lepton flavors $(\ell=e, \mu)$ in the final states, i.e. $\mu\bar{\mu}\nu\bar{\nu}$ and $e\bar{e}\nu\bar{\nu}$, which are opposite sign same flavor (OSSF) events, and $\mu \bar{e}\nu\bar{\nu}$ and $e \bar{\mu}\nu\bar{\nu}$, which are opposite sign opposite flavor (OSOF) events.  The analysis includes all possible charge combinations, as the Higgs can also decay 
into anti-heavy neutrino $(\bar{N})$ for a Dirac-type $N$ or $N$ can decay to both positively and negatively charged leptons for a Majorana-type $N$. 

To analyze the $2\ell 2\nu$ final states obtained from our detector simulation, we use the selection cuts listed below from the ATLAS analysis~\cite{ATLAS:2014aga}. 
For $\mu\bar{\mu}$ events, we impose the following cuts:  
\begin{itemize}
\item [(i)] Transverse momentum of sub-leading lepton: $p^{\ell_2, {\rm sub-leading}}_T > 10$ GeV.
\item [(ii)] Transverse momentum of leading lepton: $p^{{\ell_1},{\rm leading}}_{T} > 22$ GeV.
\item [(iii)] Jet transverse momentum: $p_T^j > 25$ GeV. 
\item [(iv)] Pseudo-rapidity of leptons: $|\eta^{\ell_{1, 2}} |< 2.4$ and of jets: $|\eta^j |< 2.4$.
\item [(v)] Lepton-lepton separation: $\Delta R_{\ell\ell} > 0.3$, lepton-jet separation: $\Delta R_{\ell j} > 0.3$ and jet-jet separation: $\Delta R_{jj} > 0.3$. 
\item [(vi)] Invariant mass of each OSSF lepton pair: $m_{\ell\ell}>12$  GeV. 
\item [(vii)] Transverse mass\footnote{$m_T=\sqrt{(E^{\ell\ell}+p_T^{\nu\nu})^2-|\vec{p_T}^{\ell\ell}+\vec{p_T}^{\nu\nu}|^2}$ where $E_{T}^{\ell\ell}= \sqrt{(p_T^{\ell\ell})^{2}+ (m_{\ell\ell})^{2}}$, where $\vec{p_{T}}^{\nu\nu}(\vec{p_{T}}^{\ell\ell})$ is the 
vector sum of the neutrino (lepton) transverse momenta, and $p_{T}^{\nu\nu}(p_T^{\ell\ell})$ is its magnitude.} $m_T$: $ \frac{3}{4}M_h < m_{T} < M_h$.  
\item [(viii)] Missing transverse energy (MET): $\slashed{E}_{T} > 40$ GeV. 
\item[(ix)] Events with missing transverse momentum are suppressed by requiring $p_T^{\rm miss}$ to point away from the dilepton transverse momentum, i. e. , $\Delta\phi^{\ell\ell, \rm MET} > \frac{\pi}{2}$. 
\item [(x)] Magnitude of dilepton momentum: $p_{T}^{\ell\ell} > 30$ GeV.
 \end{itemize}
For $e\bar{e}$ events, similar cuts are applied, except for the pseudo-rapidity of leptons: $|\eta^{\ell_{1, 2}} |< 2.47$. 
For $\mu\bar{e}(e\bar{\mu})$ events, the only differences are $|\eta^e| < 2.47, |\eta^\mu| < 2.4$, $m_{e\mu}>10$  GeV and $\slashed{E}_{T} > 20$ GeV. 

\begin{figure}[t!]
\centering
\includegraphics[scale=0.8]{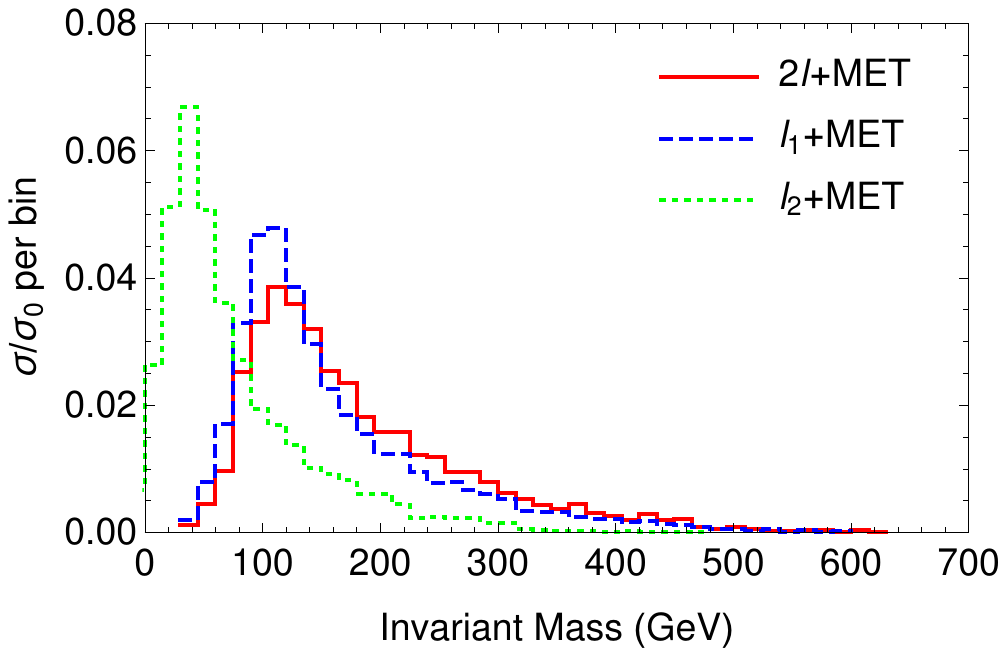}
\caption{$2\ell 2\nu$ event distributions for $M_N=100$ GeV.}
\label{fig:dist1}
\end{figure}

\begin{figure}
\begin{center}
\includegraphics[scale=0.45]{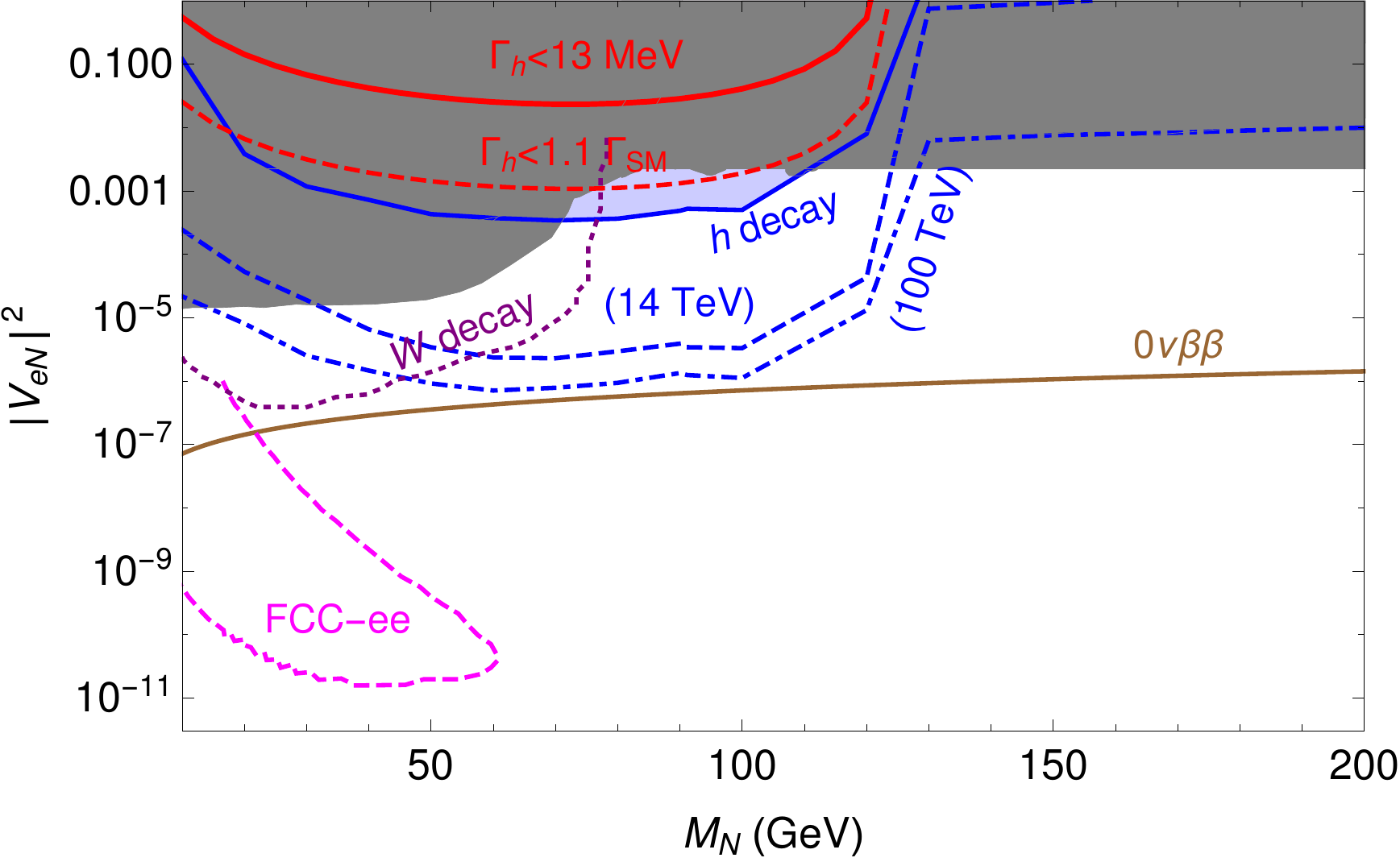}
\includegraphics[scale=0.45]{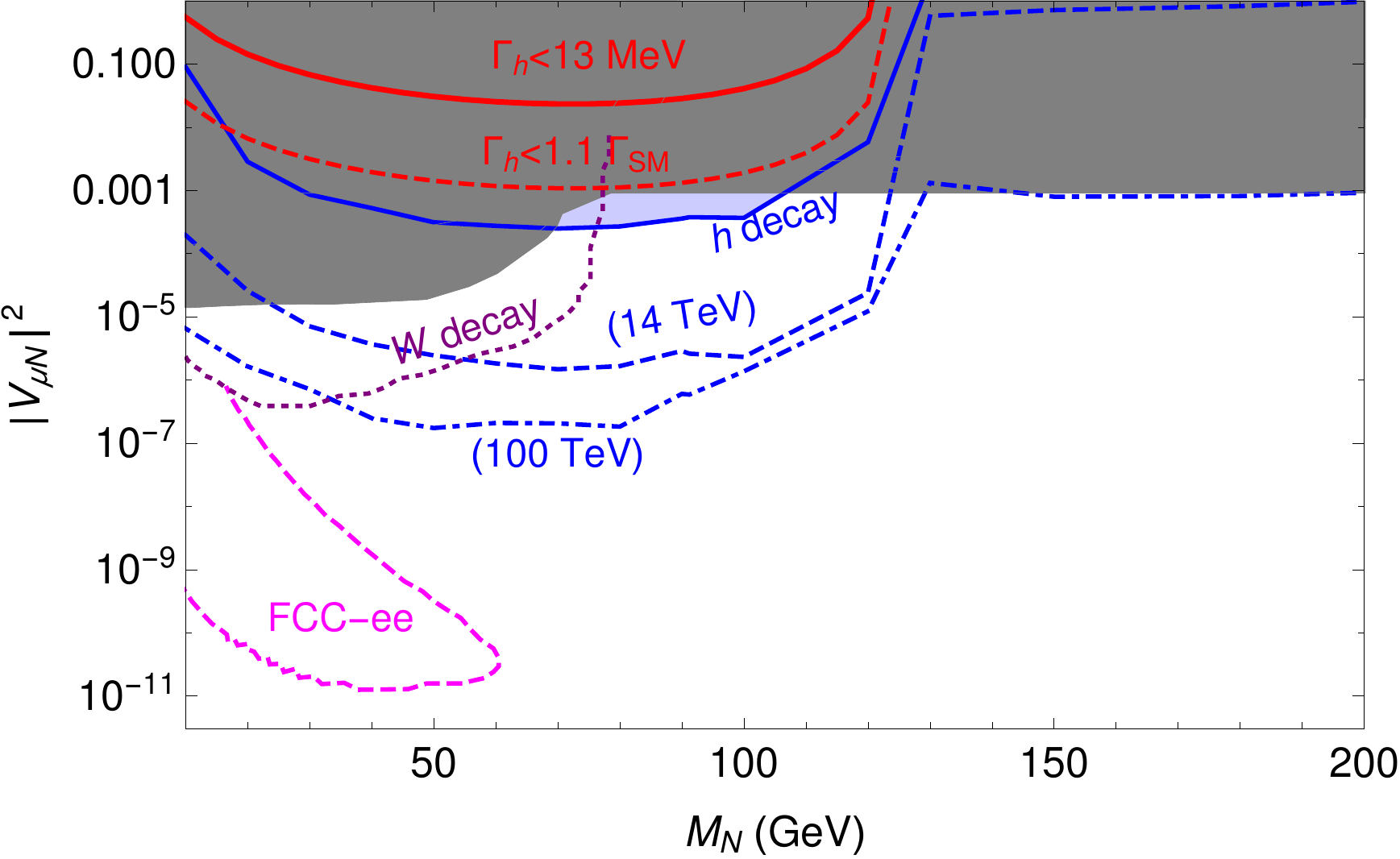} \\
\includegraphics[scale=0.45]{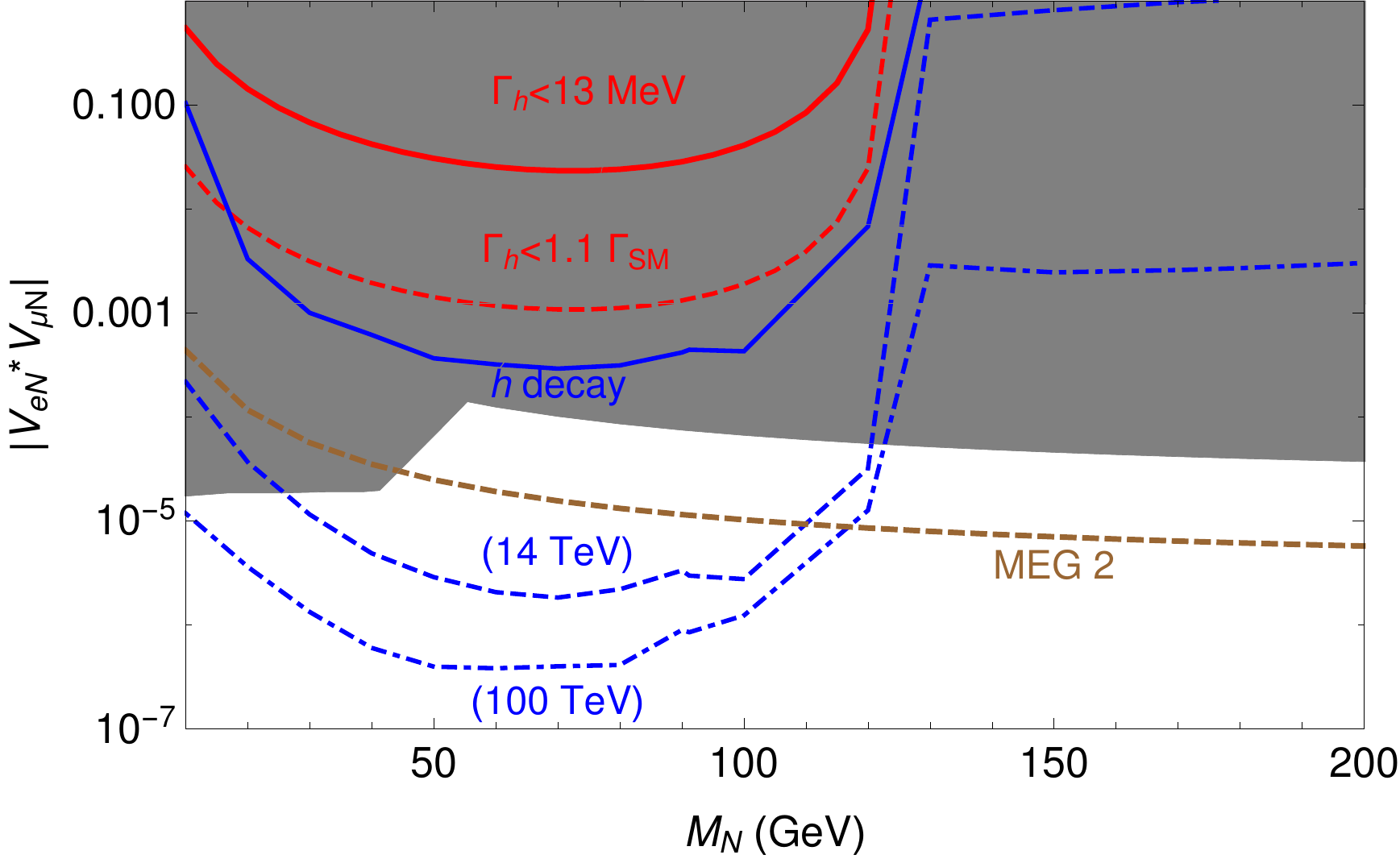}
\end{center}
\caption{Upper bound on the mixing angle from the $h\to 2\ell2\nu$ channel at the LHC. The left panel in the upper row stands for $2\mu2\nu$, the right panel shows the result for $2e2\nu$
final state, and the lower row stands for $e\mu 2\nu$ channel. The shaded regions in each panel is experimentally excluded from a combination of low and high-energy searches for sterile neutrinos. For comparison, we also show the corresponding current/future limits from a few other relevant experiments. For details, see text. }
\label{fig2}
\end{figure}

The relevant background to these final states are mainly from $WW$ (irreducible), top quarks (both single and pair produced), misidentified leptons (from $Wj$ and $jj$), other dibosons ($W\gamma$, $Z\gamma$, $WZ$, $ZZ$) and Drell-Yan processes ($Z/\gamma^*\to \ell \ell$). The distinguishing features of these backgrounds motivate the definition of the event categories based on the lepton flavor, as mentioned above.  For a detailed discussion of the background separation using specific kinematic features, see Refs.~\cite{Chatrchyan:2013iaa, ATLAS:2014aga, CMS:2016nfx, ATLAS:2016kjy}. Here we just illustrate a few relevant distributions in Fig.~\ref{fig:dist1}, namely, the invariant masses of the dilepton+MET and  lepton+MET events for a typical value of $M_N=100$ GeV. As expected, the dilepton+MET distribution peaks around the Higgs boson mass, which is one of the main features of the signal not exhibited by the background. 

After imposing the selection cuts from ATLAS listed above, we calculate the yield of events from the detector simulation for the three different final states (OSSF and OSOF) to compute the corresponding bounds on the square of the light-heavy neutrino mixing parameter as a function of the heavy neutrino mass: 
\bea
{\cal N}(M_N, |V_{\ell N}|^{2})&\ = \ & L\cdot \sigma^{\rm SM}_{h} \Big[\epsilon^{\rm SM} \frac{\Gamma(h\rightarrow WW^*\rightarrow \ell\bar{\ell}\nu \bar{\nu})}{\Gamma_{\rm SM}+ \Gamma_{\rm New}}+ 
\sum_{j, k}  \epsilon_{jk} \frac{\Gamma(h\rightarrow \bar{\nu}N+ {\rm c.  c.} \rightarrow \ell_j\bar{\ell}_k \nu \bar{\nu})}{\Gamma_{\rm SM}+ \Gamma_{\rm New}}\Big]\nonumber \\
\label{yield}
\eea
where $L$ is the is the integrated luminosity, $\sigma^{\rm SM}_{h}(pp\to h)$ is the SM Higgs production cross section (which is dominantly from the gluon-gluon fusion through a top-quark loop and not affected by the new Yukawa interaction), $j,k$ are flavor indices $e,\mu$, and $\epsilon_{\rm SM}, \epsilon_{jk}$ are the efficiencies for the decays mediated by the SM and in presence of the sterile neutrino, respectively, calculated using the selection cuts listed above. For the total width of the SM Higgs boson $\Gamma_{\rm SM}$ and the partial width 
$\Gamma(h\rightarrow WW^*\rightarrow \ell\bar{\ell} \nu \bar{\nu})$ we take the reference values given in Ref.~\cite{Heinemeyer:2013tqa} for $M_h=125$ GeV. For the production cross sections at the $\sqrt s=8$ TeV LHC, we use the reference values from Ref.~\cite{HiggsXsec1}, and for those at the 14 TeV LHC and 100 TeV hadron collider, we take the results from Ref.~\cite{HiggsXsec2}. 

To derive an upper bound on the mixing parameter, we compute the maximal value of $|V_{\ell N}|^{2}$ such that ${\cal N}(M_N, |V_{\ell N}|^{2}) < {\cal N}_{\rm expt}$, where ${\cal N}_{\rm expt}= 169$ denotes the $95\%$ CL upper limit on the number of excess $2\ell 2\nu$ events for $M_{h}= 125$ GeV at $\sqrt s=8$ TeV with $L=20.3~{\rm fb}^{-1}$~\cite{ATLAS:2014aga}. We plot this bound on the mixing parameter as a function of the $M_N$ in Fig.~\ref{fig2} (blue solid curves) for three different cases, depending on whether the $N$ mass eigenstate only couples to the electron flavor (top left panel), muon flavor (top right panel) or both (bottom panel). Assuming the same ${\cal N}_{\rm expt}$ for $\sqrt s=14$ and 100 TeV colliders, but with an integrated luminosity of 3000 fb$^{-1}$, we also show the corresponding future limits (blue dashed and dot-dashed curves, respectively).

For comparison, we also show in Fig.~\ref{fig2} various other constraints from both low and high-energy searches for sterile neutrinos. The shaded region is excluded from a combination of the LEP, LHC and electroweak precision data, and lepton flavor violation (LFV). For a detailed discussion of these constraints, see e.g. Refs.~\cite{Atre:2009rg, Deppisch:2015qwa, Antusch:2014woa, deGouvea:2015euy, Fernandez-Martinez:2016lgt, Lopez-Pavon:2015cga, Drewes:2016jae, Alonso:2012ji} and references therein.  The future limits from $W$ decay at $\sqrt s=14$ TeV LHC~\cite{Izaguirre:2015pga} and $Z$ decay at FCC-ee~\cite{Blondel:2014bra} are also shown. For the electron flavor, the most stringent limit is obtained from the non-observation of $0\nu\beta\beta$~\cite{KamLAND-Zen:2016pfg, Agostini:2017jkl}, as shown by the brown solid curve in the top left panel of Fig.~\ref{fig2}. For deriving this limit, we have assumed the heavy neutrino to be Majorana and dominantly contributing to $0\nu\beta\beta$~\cite{Mitra:2011qr}. For (pseudo) Dirac neutrinos, this limit does not apply. Similarly, the bob-observation of LFV processes such as $\mu\to e\gamma$~\cite{MEG:2016wtm} put stringent constraints on the mixing combination $V_{eN}^*V_{\mu N}$, and the future MEG 2 upgrade can improve this limit significantly, as shown in the bottom panel of Fig.~\ref{fig2}. 
Here we have also included the LFV limits from direct heavy neutrino searches at CMS~\cite{Khachatryan:2016olu}. 

We find that the limits derived from Higgs decay are the strongest when $M_N$ is in the vicinity, but below the Higgs mass. The limits derived from $\sqrt s=8$ TeV LHC Higgs data are better than the current global constraints on sterile neutrinos in the mass range 70-110 GeV for $|V_{\ell N}|^2$, whereas for $V_{eN}^*V_{\mu N}$, the MEG limit is still the most stringent one. The Higgs decay limits become ineffective as $M_N$ approaches $M_h$ for kinematic reasons. Nevertheless, with more precision Higgs measurements in the near future, the limits derived from the Higgs decay could be improved substantially.

\section{Sterile neutrino production with $\ell \nu jj$ final state}
\label{sec:calc}

If the $W$ boson produced in the Higgs decay to $\nu N\to \nu \ell W$ decays hadronically, it will give rise to $\ell \nu jj$ final state,  which is complementary to the $2\ll 2\nu$ channel discussed in the last section. Since the hadronic branching ratio of $W$ (67\%) is almost three times the leptonic branching ratio (22\%, for $e,\mu$ combined), the $\ell \nu jj$ final state is supposed to give a larger signal cross section at the LHC. However, the pure leptonic modes are much cleaner in the hadron collider environment, whereas the $\ell \nu jj$ channel suffers from a much larger irreducible background, mostly from $WW$ and $WZ$.  Thus, it turns out that the signal sensitivity in the $\ell \nu jj$ channel is smaller than the $2\ell 2\nu$ channel. Nevertheless, due to the presence of only one neutrino in the final state, the event reconstruction is easier in this case. So this section is devoted to the discussion of this channel. 

\begin{figure}[t!]
\begin{center}
\includegraphics[scale=1]{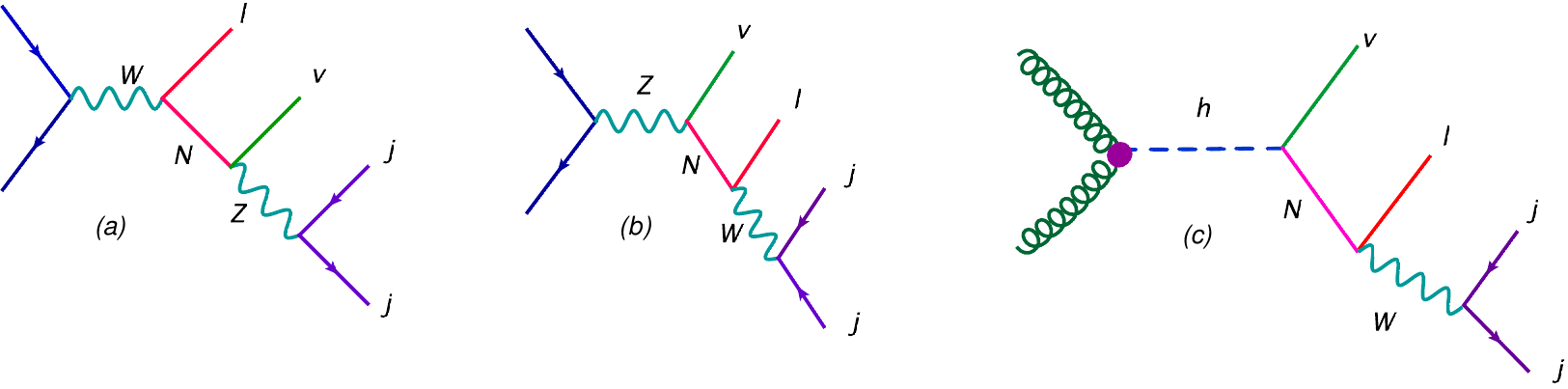}
\end{center}
\caption{$\ell\nu jj $ final state from heavy neutrino production and decay at the hadron colliders. }
\label{fig5}
\end{figure}

Apart from its production from Higgs decay mediated by the Dirac Yukawa coupling \eqref{yuk}, the heavy neutrino can also be produced at colliders through the CC interaction in Eq.~\eqref{CC} and the NC interaction in Eq.~\eqref{NC}, which in turn could contribute to the $\ell \nu jj$ channel, as shown in Fig.~\ref{fig5}. We include all these processes in our analysis of the $\ell \nu jj$ signal. 

We use the event generator  {\tt MadGraph5-aMC@NLO}~\cite{aMC}  to produce the events at parton level and perform the  
showering and hadronization of the events with {\tt PYTHIA6.4}~\cite{Pyth} bundled in {\tt MadGraph} with {\tt anti-$k_{T}$} algorithm, while the jets are 
clustered using {\tt FastJet}~\cite{FJ}. To calculate the hadronic cross-sections we use the {\tt CTEQ6L1} PDF~\cite{Dulat:2015mca}. The hadronized events are passed through {\tt Delphes}~\cite{Delphes} to simulate the detector response. 

\begin{figure}[t!]
\centering
\includegraphics[scale=0.7]{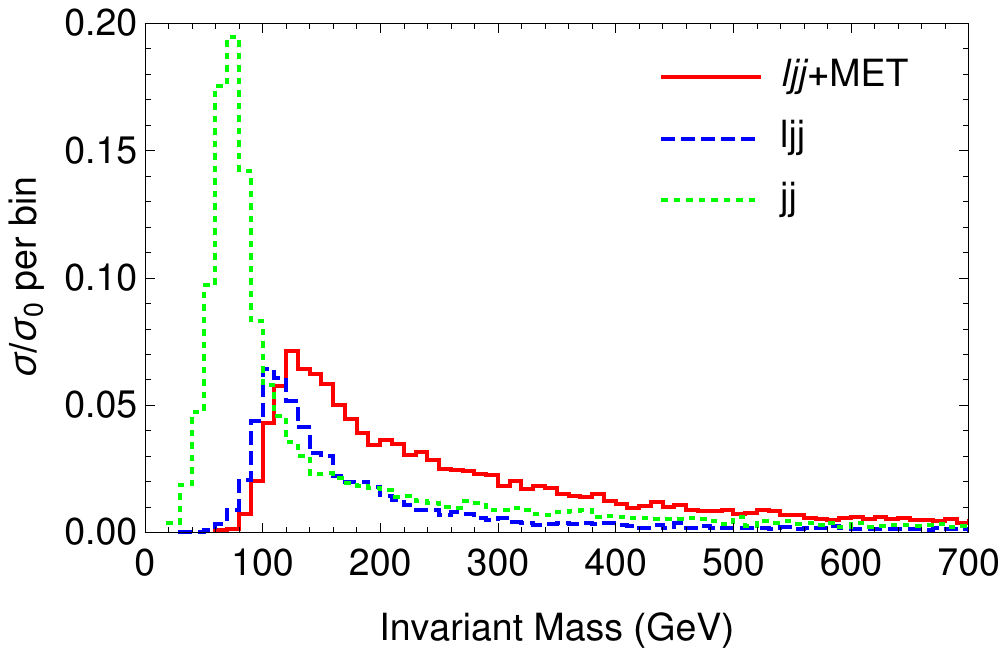}
\includegraphics[scale=0.7]{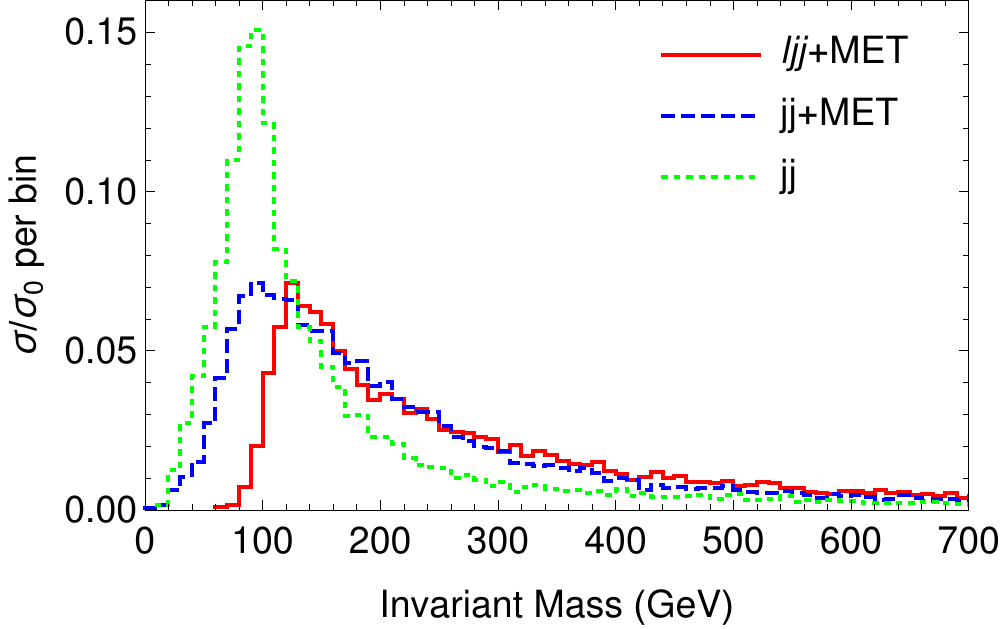}
\caption{$\ell \nu jj$ invariant mass distributions for $M_N=100$ GeV. The left panel corresponds to the $W\to jj$ final state, whereas the right panel corresponds to $Z\to jj$ final state. } \label{fig7}
\end{figure}

The selection cuts used in our analysis for optimizing the signal-to-background  are listed below for different center-of-mass energies. For $\sqrt s=8$ TeV, we have imposed the following cuts: 
\begin{itemize}
\item [(i)] Transverse momentum of the lepton: $p^{\ell}_T > 20$ GeV.
\item [(ii)] Transverse momentum of jets: $p^{j_{1,2}}_T > 30$ GeV.
\item [(iii)] Pseudo rapidity of lepton: $|\eta_{\ell}| < 2.5$.
\item [(iv)] Pseudo-rapidity of jets: $|\eta^{j_{1, 2}}| < 2.5$.
\item[(v)] Lepton-jet separation $\Delta R_{\ell j} > 0.3$ and jet-jet separation $\Delta_{jj} > 0.4$.
\item[(vi)] Invariant mass cut for the reconstruction of the of the heavy neutrino and the gauge boson produced after the heavy neutrino decay: $m_i -20 < m_i < m_i +20$, 
where $m_i= M_{N}, m_{W}$ or $m_{Z}$ depending on the processes given by the Feynman diagrams in Fig.~\ref{fig5}. To reconstruct $M_N$ we use the invariant mass $m_{\nu jj}$ for Fig.~\ref{fig5}(a) and $m_{\ell jj}$ from Figs.~\ref{fig5}(b) and (c).  The SM gauge bosons are reconstructed from the invariant mass $m_{jj}$. The various invariant mass distributions are shown in Fig.~\ref{fig7} for a typical choice $M_N=100$ GeV for illustration. 
\end{itemize}
For $\sqrt s= 14$ TeV, we use the same selection cuts, except for $p^{\ell}_T > 30$ GeV and $p^{j_{1,2}}_T > 32$ GeV. For $\sqrt s=100$ TeV, we use even stronger cuts: $p^{\ell}_T > 53$ GeV and $p^{j_{1,2}}_T > 35$ GeV, while the other cuts remain the same as in the 8 TeV case. 
Our analysis is done for the $\ell =\mu$ case only, which gives better sensitivity than the $\ell=e$ case. 

\begin{figure}
\begin{center}
\includegraphics[scale=0.5]{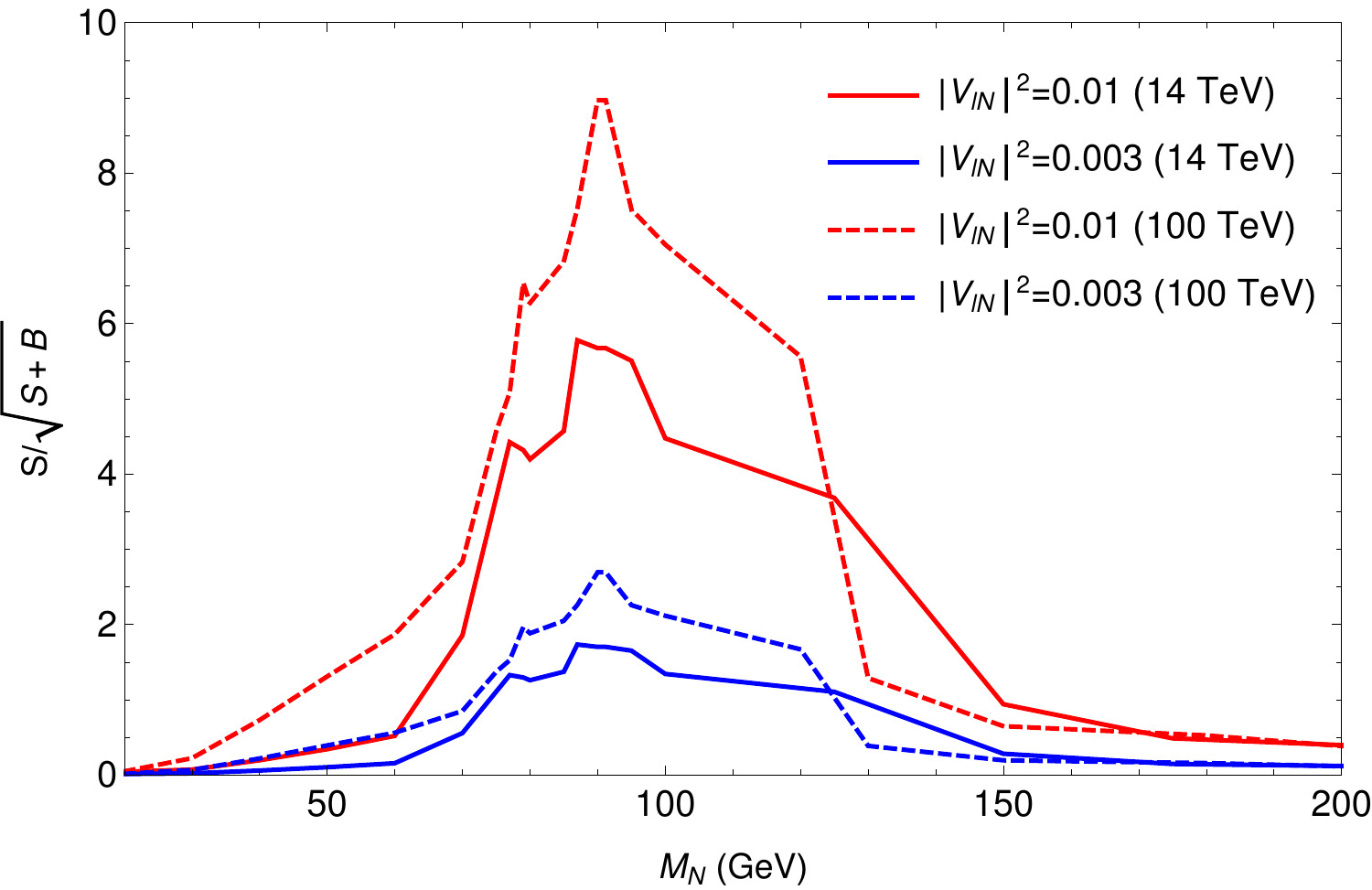}
\end{center}
\caption{Significance of the $\ell\nu jj $ final state at $\sqrt s=14$ and 100 TeV for two different choices of $|V_{\ell N}|^2$. }
\label{fig6}
\end{figure}

For the dominant SM background, we have considered the irreducible backgrounds from the $WW$ and $WZ$ processes. 
After examining the signal  ($S$) and background ($B$) efficiencies, we calculate the significance of the $\ell \nu jj$ channel, defined as 
\bea
{\cal N} =\frac{S}{\sqrt{S+B}}
\label{sig}
\eea
where $S\propto |V_{\ell N}|^{2}$.  Our combined results for the three channels shown in Fig.~\ref{fig5} are given in  Fig.~\ref{fig6} as a function of the heavy neutrino mass for two different choices of $|V_{\ell N}|^2=0.01$ (red) and 0.003 (blue) and for $\sqrt s= 14$ TeV (solid) and 100 TeV (dashed) with integrated luminosity of 3000 fb$^{-1}$. The results for the $\sqrt s=8$ TeV case are not so promising and hence not shown here. 

We find that for $|V_{\ell N}|^{2}=0.01$ (at the edge of the current upper limit),  the $\ell \nu jj$ channel has more than 3$\sigma$ significance in the mass range $M_N=70-120$ GeV. For smaller $|V_{\ell N}|^{2}$, the signal sensitivity decreases rapidly and for $|V_{\ell N}|^{2}=0.003$, it cannot reach 3$\sigma$ for any mass value. Going to $\sqrt s=100$ TeV increases the significance in the same mass range, but drops rapidly on either side of this mass range. 


\section{Conclusion}\label{sec:conc}
We have studied the sterile neutrino production in Higgs decays mediated by the Dirac Yukawa coupling in the singlet seesaw extension of the SM. This Yukawa coupling, which is responsible for the light neutrino masses in the seesaw mechanism, also induces the Higgs decay $h\to \nu N$, thus affecting its total decay width, as well as its partial widths in certain channels, $WW^*$ in particular. Using the $\sqrt s=8$ TeV LHC Higgs data in the $WW^*\to 2\ell 2\nu$ channel, we derive stringent constraints on the active-sterile neutrino mixing parameter in the sterile neutrino mass range close to the Higgs mass. With precision Higgs measurements in the near future, we expect these limits to further improve significantly.

We have also studied a new final state for the heavy neutrino production, namely, $\ell \nu jj$ from the Higgs and $W,~Z$ mediated processes. It turns out that the signal sensitivity in this channel is smaller than the $2\ell 2\nu$ channel, but due to the presence of only one neutrino in the final state, it offers the possibility of a better signal reconstruction. We find that a $3\sigma$ significance in the $\ell \nu jj$ channel is possible for sterile neutrino masses in the mass range between 70 and 120 GeV.  
\bigskip
\acknowledgments
The work A.D. is supported by the Korea Neutrino Research Center which is established by the National Research Foundation of Korea (NRF) grant funded by the Korea government (MSIP) (No. 2009-0083526).
The work of C.S.K. is supported by the NRF grant funded by the Korean government
of the MEST (No. 2016R1D1A1A02936965).

\end{document}